\documentclass[11pt]{JHEP3}

\newcommand{\vt}{{{\vartheta}}}

\newcommand{\N}{{\cal N}}

\newcommand{\be}{\begin{equation} }
\newcommand{\ee}{\end{equation} }
\newcommand{\ba}{\begin{array}}
\newcommand{\ea}{\end{array}}

\newcommand{\bea}{\begin{eqnarray}}
\newcommand{\eea}{\end{eqnarray}}

\def\O{{\cal O}}

\def\S{{\cal S}}

\def\tr{{\rm tr}}

\title{Superfield formalism for the one loop effective action and CP($N$) 
model in three dimensions}

\author{Jin-Ho Cho, Sang-Ok Hahn, Phillial Oh, Cheonsoo Park and Jeong-Hyuck
Park${}^{\dagger}$\\
\\
 Department of Physics and Institute of Basic Science, Sungkyunkwan University\\
 Chunchun-dong, Jangan-gu, Suwon 440-746, Korea\\
\\
${}^{\dagger}$Institut des Hautes Etudes Scientifiques, Bures-sur-Yvette,
F-91440, France \\
}
\abstract{To obtain the one loop effective action for a given
superfield theory, one encounters  the notion such as the
`supertrace' of a differential operator on superspace. We
develop,  in a systematic way for the superspace of arbitrary
dimension, a  method to determine the supertrace precisely.
We present a formula to express the supertrace explicitly  as the
superspace integral, which enables us to write the one loop
effective action within the superfield formalism and still
maintain the manifest supersymmetry. In the second part of the
paper, we apply the result to a three dimensional  $\N=1$
supersymmetric CP($N$) model in the auxiliary superfield
formalism. The model  contains a novel topological
interaction term. We show in the large $N$ limit the one loop
effective action is given by  the supersymmetric
Maxwell-Chern-Simons theory. }


\preprint{hep-th/0312088}

\begin{document}

%%%%%%%%%%%%%%%%%%%%%%%%%%%%%%%%%%%%%%%%%%%%%%%%%%%%%%%%%%%%%%%%%%%%%%%%%%%%%%%%%%%%%%%%%%%%%%%%%%%%%%%%%%%%%%%%%%%%%%%%%%%%%%%%%%%%%%%%%%%%%%%%%%%%%%%%%%%%%%%%%%%%%%%%%%%%%%%%%%%%%%%%%%%%%%%%%%%%%%%%%%%%%%%%%%%%%%%%%%%%%%%%%%%%%%%%%%%%%

\section{Introduction and summary}
There are two approaches to the path integral quantization of
supersymmetric field theories. One is to work in the component
formalism. The functional integral is straightforward, but
supersymmetry is not manifest. The other is to go to the
superspace \cite{sala}  which contains the anti-commuting
Grassmann coordinates as well as the usual spacetime coordinates.
The ordinary fields and their functional integrals are replaced
by the superfields and super functional integrals \cite{bere}. In
consequence, we encounter the notions  such as superdeterminant
and supertrace, and these quantities are to be directly evaluated
on superspace. Superpropagators and super Feynman rules naturally
follow on superspace and the supersymmetric effective action can
be
\textit{canonically} computed while keeping the supersymmetry manifest \cite{wess,gate}.\\

In the present paper, for the superspace of arbitrary dimension,
we develop a novel  technique to determine precisely the
supertrace of a differential operator on superspace, which arises
in the computation of the one loop effective action
\textit{within the path integral formalism}. We present a formula
to express the supertrace explicitly  as the superspace integral,
which enables us to write the one loop effective action within
the superfield formalism and still maintains the manifest
supersymmetry. In the second part of the paper, we apply the
result  to a three dimensional $\N=1$ supersymmetric CP($N$)
model in the auxiliary gauge superfield formalism
\cite{golo,aref,aoya}. The model also contains   a topological
interaction term.  Its bosonic sector is the higher derivative
CP($N$) model with the Wess-Zumino-Witten term and also the
topological current term squared \cite{itoh}. Since the theory has
the interesting properties in the large $N$ limit such as  its
renormalizability and the Maxwell-Chern-Simons terms being
dynamically induced at a nontrivial UV fixed point, it is
worthwhile to check how the properties persist in the supersymmetric case too.\\

The organization of the paper is as follows. Section
\ref{general} contains our main result for the  one loop
effective action on the superspace of arbitrary dimension.  We
consider an arbitrary differential operator on the superspace,
$\Delta$,  which generally depends on the superspace coordinates
as well as their derivatives. The one loop effective action then
corresponds to the ``supertrace of the logarithm of it''. We
conceive a certain dual orthonormal basis for the Grassmannian
coordinates and express the differential operator as well as the
superfields of the given theory in terms of these basis. The
differential operator then corresponds to a supermatrix,
$\tilde{\Delta}$, while the superfields are matched to the
$Z_{2}$-graded vectors on which the supermatrix acts. Moreover,
the computation of the one loop effective action within the path
integral formalism yields the supertrace of the logarithm of the
supermatrix corresponding to
 the differential operator, $\mbox{Str}(\ln\widetilde{\Delta})$.
Following the very definition of the supertrace for the
supermatrix, we explicitly compute this quantity. Firstly, thanks
to the orthonormal  property of the dual basis, the supermatrix
corresponding to the product of two differential operators is
identical to the product of the two corresponding supermatrices,
and hence we obtain the crucial identity,
$\mbox{Str}(\ln\widetilde{\Delta})=\mbox{Str}(\!\widetilde{~\ln{\Delta}~}\!)$.
Secondly, we  demonstrate how to manipulate the supertrace of the
supermatrix corresponding to an operator on superspace  in terms
of the superspace integral with the operator itself rather than
the supermatrix. Combining the two results, we  obtain the final
formula  which expresses $\mbox{Str}(\ln\widetilde{\Delta})$ as
the superspace integral of the multi-commutators between the
logarithm of the operator, $\ln\Delta$, and the Grassmann
coordinates.  Our result can be rewritten in an alternative
fashion to resemble some known conventional expression for the
supertrace. We believe our result clarifies  the precise meaning
of it.

In section \ref{secCP}, we apply the above result  to the three
dimensional U(1) gauged $\N=1$ supersymmetric CP($N$) model
which  contains the gauge superfield and also a topological
interaction term.  Sticking to the superspace formalism, we solve
the mass gap equation and compute the one-loop effective action.
We  show that in the large $N$ limit, the theory is
renormalizable and the one loop effective action is given by the
supersymmetric Maxwell-Chern-Simons theory plus
its higher  spacetime derivative generalization.
In particular, the first order in the derivative expansion corresponds to the
supersymmetric generalization of the
extended topological massive electrodynamics model studied by  Deser and Jackiw~\cite{jack}.

The Appendix demonstrates  the relation of  the gauged CP($N$)
model to the supersymmetric higher derivative CP($N$) model with
the Wess-Zumino-Witten term~\cite{itoh}.

%%%%%%%%%%%%%%%%%%%%%%%%%%%%%%%%%%%%%%%%%%%%%%%%%%%%%%%%%%%%%%%%%%%%%%%%%%%%%%%%%%%%%%%%%%%%%%%%%%%%%%%%
%%%%%%%%%%%%%%%%%%%%%%%%%%%%%%%%%%%%%%%%%%%%%%%%%%%%%%%%%%%%%%%%%%%%%%%%%%%%%%%%%%%%%%%%%%%%%%%%%%%%%%%%
%%%%%%%%%%%%%%%%%%%%%%%%%%%%%%%%%%%%%%%%%%%%%%%%%%%%%%%%%%%%%%%%%%%%%%%%%%%%%%%%%%%%%%%%%%%%%%%%%%%%%%%%
\section{One loop effective action - general analysis\label{general}}
We consider a generic superspace of arbitrary dimension, $d+N_{\!f}$, where
the even part is a spacetime of dimension, $d$, and the odd part consists of
Grassmann variables, $\vartheta^{\alpha}$, $\alpha=1,2,\cdots,N_{\! f}$. Without loss of generality  we take
the real basis so that $\vartheta^{\alpha}=(\vartheta^{\alpha})^{\dagger}$ and let $N_{\! f}$ be a even number.

On the superspace the superfield is of the generic form,
\begin{equation}
\Phi(z)=\phi(x)+\vartheta^{\alpha}\psi_{\alpha}(x)+\cdots+{\vartheta^{1}
\vt^{2}\cdots\vt^{N_{\!f}}}F(x)=\displaystyle{\sum_{n=0}^{N_{\!f}}\,
\frac{1}{n!}{\vt^{\alpha_{1}}\vt^{\alpha_{2}}\cdots\vt^{\alpha_{n}}}
\phi_{\alpha_{1}\alpha_{2}\cdots\alpha_{n}}(x)\,.} \label{grass1}
\end{equation}
Its complex conjugate superfield reads
\be
\Phi^{\dagger}=\displaystyle{
\sum_{n=0}^{N_{\!f}}\,\frac{1}{n!}{\phi_{\alpha_{1}\alpha_{2}
\cdots\alpha_{n}}^{\dagger}
\vt^{\alpha_{n}}\cdots\vt^{\alpha_{2}}\vt^{\alpha_{1}}}\,.}
\ee

Any operator acting on the superfield is a function of the superspace
coordinates and  their derivatives,
$\Delta(x^{\mu},\vartheta^{\alpha},\partial_{\nu},\partial_{\beta})$.
The typical manipulation to get the one loop effective action  is to integrate
out the complex scalar superfields,
\be
\displaystyle{\int D\Phi D{\Phi^{\dagger}}\,e^{\int\!dz^{d+N_{\! f}}\,
{\Phi^{\dagger}}\Delta\Phi}=\frac{1}{``\mbox{sdet}\Delta"}
=e^{-``\mbox{Str}(\ln \Delta)"}}\,.
\label{oneloop}
\ee

In the remaining of this section we show that the precise meaning of
$``\mbox{Str}(\ln \Delta)"$ is in fact,
\be
\ba{ll}
``\mbox{Str}(\ln \Delta)"&=\displaystyle{\int \!dz^{d+N_{\!f}}\,
\frac{1}{N_{\!f}!}\,\,
\epsilon^{\alpha_{1}\alpha_{2}\cdots\alpha_{N_{\!f}}}
\langle x|\{[\cdots\{[\ln\Delta,\vt^{\alpha_{1}}],\vt^{\alpha_{2}}\}
\cdots]\vt^{\alpha_{N_{\!f}}}\}|x\rangle}\\
{}&{}\\
{}&=\displaystyle{\int \!dz^{d+N_{\!f}}\,
\lim_{z^{\prime}\rightarrow z}\,\langle z^{\prime}|\ln\Delta|z\rangle
=\int \!dz^{d+N_{\!f}}\,
\lim_{\vt^{\prime}\rightarrow \vt}\,\langle x|\ln\Delta
\delta(\vt-\vt^{\prime})|x\rangle\,,}
\ea
\label{str}
\ee
where $\delta(\vt-\vt^{\prime})=(\vt^{1}-\vt^{\prime 1})
(\vt^{2}-\vt^{\prime 2})\cdots(\vt^{N_{\!f}}
-\vt^{\prime \N_{\!f}})$.
The second line resembles some known result in the literature \cite{witt},
where the supertrace is simply given by
$\int\!dz^{d+N_{\!f}}\,\langle z|\O|z \rangle$. Hence, our result clarifies
the precise meaning of the conventional expression.\\

\textit{Proof}\\
We first let all the possible independent products of the Grassmann
coordinates, \\
$\{{\vt^{\alpha_{1}}\vt^{\alpha_{2}}\cdots\vt^{\alpha_{n}}},~n=0,1,
\cdots,N_{f}\}
\equiv\{\,|M\rangle\,\}$ be a basis for the $\vt^{\alpha}$
expansion of the superfield with the dimension, $2^{N_{\!f}}$,  i.e.
$M=1,2,\cdots,2^{N_{\!f}}$.  Then formally  we can put
\begin{equation}
\Phi(z)=\sum_{M}\,|M\rangle\, \widetilde{\Phi}^{M}(x)\,, \label{grass2}
\end{equation}
and write an associated  column vector,
\begin{equation}
\widetilde{\Phi}=(\widetilde{\Phi}^{1},\widetilde{\Phi}^{2},\cdots,
\widetilde{\Phi}^{M},\cdots\,)^{T}\,.
\end{equation}

Now we define the dual orthonormal basis,
$\{\,\langle K|\,\}$ by $\langle K|M\rangle=\delta_{KM}$.
Explicitly it is of the form
\be
\langle K|=\partial_{1}\partial_{2}\cdots\partial_{N_{\!f}}
\vartheta^{N_{\!f}}\cdots\vt^{2}\vt^{1}
\partial_{\beta_{m}}\cdots\partial_{\beta_{2}}\partial_{\beta_{1}}\,,
\label{grass3}
\ee
and the orthonormality reads
\be
(\partial_{1}\partial_{2}\cdots\partial_{N_{\!f}}\vartheta^{N_{\!f}}
\cdots\vt^{2}\vt^{1}
\partial_{\beta_{m}}\cdots\partial_{\beta_{2}}\partial_{\beta_{1}})
(\vt^{\alpha_{1}}\vt^{\alpha_{2}}\cdots\vt^{\alpha_{n}})=
n!\,\delta_{m}^{~n}\,\delta^{~[\alpha_{1}}_{\beta_{1}}
\delta^{~\alpha_{2}}_{\beta_{2}}\cdots\delta^{~\alpha_{n}]}_{\beta_{n}}\,,
\ee
where on the right hand side $[\alpha_{1}\alpha_{2}\cdots\alpha_{n}]$
denotes the total anti-symmetrization having the ``strength one" or the
multiplication factor, $1/{n!}\,$. The usual completeness relation follows straightforwardly,
\be
\sum_{K}\,|K\rangle\langle K|=\mbox{Identity}\,.
\ee

Using the orthonormal basis above  we can express any operator acting on the
superfields,
\be
\O(x^{\mu},\vt^{\alpha},\partial_{\nu},\partial_{\beta})=
|M \rangle\widetilde{\O}_{MK}(x^{\mu},\partial_{\nu})\langle K|\,.
\ee

The crucial merit of conceiving the above orthonormal dual basis is as follows.
The product of operators can be also rewritten in terms of the basis.
From the orthonormality we get
\be
\ba{ll}
\O_{1}\O_{2}=|M \rangle\widetilde{~\O_{1}\O_{2}~}_{MK}\langle K|\,,~~~&~~~
\widetilde{~\O_{1}\O_{2}~}_{MK}=\langle M| \O_{1}\O_{2}|K \rangle
=\widetilde{\O_{1}}_{ML}\widetilde{\O_{2}}_{LK}\,.
\ea\label{rep}
\ee
Namely, the supermatrix, $\widetilde{\O}_{MK}=\langle M| \O|K \rangle$,
gives a good representation for the product of the operators.

 The supertrace of the supermatrix, $\widetilde{\O}$, can be now expressed as
a superspace integration of the operator itself,
\be
\ba{ll}
\mbox{Str}\widetilde{\O}&={\displaystyle{\sum_{M}\,(-1)^{\#(M)}
\langle M|\O|M\rangle}}\\
{}&{}\\
{}&=\tr\!\left[\displaystyle{\sum_{n=0}^{N_{\!f}}\,\frac{(-1)^{n}}{n!}}\,
\partial_{1}\partial_{2}\cdots\partial_{N_{\!f}}\vartheta^{N_{\!f}}
\cdots\vt^{2}\vt^{1}
\partial_{\alpha_{n}}\cdots\partial_{\alpha_{2}}\partial_{\alpha_{1}}\O\,
\vt^{\alpha_{1}}\vt^{\alpha_{2}}
\cdots\vt^{\alpha_{n}}\right]\\
{}&{}\\
{}&=\displaystyle{\int\!d\vt^{N_{\!f}}\,\tr\!\left[\sum_{n+m=N_{\!f}}
\frac{(-1)^{n}}{\,n!\,m!\,}\,\,
\epsilon^{\alpha_{1}\alpha_{2}\cdots\alpha_{n}\beta_{1}\beta_{2}
\cdots\beta_{m}}\vt^{\alpha_{1}}\vt^{\alpha_{2}}
\cdots\vt^{\alpha_{n}}\O\,\vt^{\beta_{1}}\vt^{\beta_{2}}\cdots\vt^{\beta_{m}}
\right]}\\
{}&{}\\
{}&=\displaystyle{\int
\!d\vt^{N_{\!f}}\,\tr\!\left[\frac{1}{N_{\!f}!}\,\,
\epsilon^{\alpha_{1}\alpha_{2}\alpha_{3}
\cdots\alpha_{N_{\!f}}}\{[\cdots[\{[\O,\vt^{\alpha_{1}}],\vt^{\alpha_{2}}\},
\vt^{\alpha_{3}}]\cdots]\vt^{\alpha_{N_{\!f}}}\}\right]}\,. \ea
\label{form1}
\ee
Here  `$\tr$' denotes the trace over the spacetime coordinates,
$\epsilon^{\alpha_{1}\alpha_{2}\cdots\alpha_{N_{\!f}}}$ is the totally
anti-symmetric tensor such that  $\epsilon^{12\cdots N_{f}}=1$, and  our
convention for the integration over the Grassmann coordinates is
 $\int d\vt^{N_{\! f}}\,\vt^{1}\vt^{2}\cdots\vt^{N_{\!f}}=1$.\\

We also have
\be
\O\Phi=\sum_{M,K}\,|M \rangle\widetilde{\O}_{MK}\widetilde{\Phi}^{K}\,,
\ee
and
\be
\int d\vt^{N_{\! f}} {\Phi}^{\dagger}\Psi=\displaystyle{\sum_{n+m=N_{\!f}}\,
\frac{1}{n!m!}\,\phi_{\alpha_{1}\alpha_{2}\cdots\alpha_{n}}^{\dagger}
\epsilon^{\alpha_{n}\cdots\alpha_{2}\alpha_{1}\beta_{1}\beta_{2}
\cdots\beta_{m}}
\psi_{\beta_{1}\beta_{2}\cdots\beta_{m}}\equiv \widetilde{\Phi}^{\dagger}
\widetilde{{\cal E}}\widetilde{\Psi}}\,.
\ee

Now  the one loop effective action (\ref{oneloop}) reads, from (\ref{rep}) and
$\mbox{sdet}\,\widetilde{{\cal E}}=1$,
\be
\ba{l}
\displaystyle{\ln\left(\int D\Phi D{\Phi^{\dagger}}\,e^{\int\!dz^{d+N_{\! f}}\,
{\Phi^{\dagger}}\Delta\Phi}\right)}=
\displaystyle{-\ln\left[{\mbox{sdet}(\widetilde{{\cal E}}\widetilde{\Delta})}
\right]}=
\displaystyle{-\mbox{Str}(\ln \widetilde{\Delta})}=
\displaystyle{-\mbox{Str}(\widetilde{\,\,\ln \Delta\,})}\\
{}\\
~~~~~~~~~~~=-\displaystyle{\int \!dz^{d+N_{\!f}}\,\frac{1}{N_{\!f}!}\,\,
\epsilon^{\alpha_{1}\alpha_{2}\cdots\alpha_{N_{\!f}}}
\langle x|\{[\cdots\{[\ln\Delta,\vt^{\alpha_{1}}],\vt^{\alpha_{2}}\}
\cdots]\vt^{\alpha_{N_{\!f}}}\}|x\rangle}\,.
\ea
\ee

Instead of using the dual orthonormal basis, if we consider the expansion of
an operator by the ordinary derivatives for the Grassmann coordinates,
\begin{equation}
\displaystyle{
\O(z,\partial_{\mu},\partial_{\alpha})=\sum_{n=0}^{N_{\!f}}\,\frac{1}{n!}\,
\O(z,\partial_{\mu})^{\alpha_{1}\alpha_{2}\cdots\alpha_{n}}
\partial_{\alpha_{1}}\partial_{\alpha_{2}}
\cdots\partial_{\alpha_{n}}\,,}
\end{equation}
it is straightforward to check that \textit{only the highest order component
contributes to the supertrace.} Thus, formally introducing $|\vt\rangle$ such that
\be
\langle \vt^{\prime}|\vt\rangle=\delta(\vt-\vt^{\prime})
=(\vt^{1}-\vt^{\prime 1})(\vt^{2}-\vt^{\prime 2})
\cdots(\vt^{N_{\!f}}-\vt^{\prime \N_{\!f}})\,,
\ee
we  get
\be
\displaystyle{
\mbox{Str}\widetilde{\O}
=\int \!dz^{d+N_{\!f}}\,\langle x|\O(z,\partial_{\mu})^{N_{\!f}\cdots 21}|x
\rangle=\int
\!dz^{d+N_{\!f}}\,
\lim_{z^{\prime}\rightarrow z}\,\langle z^{\prime}|\O|z \rangle\,.}
\ee
This completes our proof.

Recently one of the authors showed that every superfield theory
can be described by a dual supermatrix model \cite{Park:2003ku}.
It will be  interesting to see how our result persists   in
the supermatrix side.

%%%%%%%%%%%%%%%%%%%%%%%%%%%%%%%%%%%%%%%%%%%%%%%%%%%%%%%%%%%%%%%%%%%%%%%%%%%%%%%%%%%%%%%%%%%%%%%%%%%%%%%%%%%%%%%%%%%%%%%%%%%%%%%%%%%%%%%%%%%%%%%%%%%%%%%%%
%%%%%%%%%%%%%%%%%%%%%%%%%%%%%%%%%%%%%%%%%%%%%%%%%%%%%%%%%%%%%%%%%%%%%%%%%%%%%%%%%%%%%%%%%%%%%%%%%%%%%%%%%%%%%%%%%%%%%%%%%%%%%%%%
\section{$\N=1$ supersymmetric CP($N$) model in three dimensions\label{secCP}}
In this section, we consider a three dimensional supersymmetric
$\N=1$ CP($N$) model with a topological interaction term. Our
notation is as follows.  The spacetime metric is
$\eta=\mbox{diag}(+--)$, and the gamma matrices are all imaginary
given by the Pauli matrices, $\gamma^{0}=\sigma^{2}$,
$\gamma^{1}=i\sigma^{3}$, $\gamma^{3}=i\sigma^{1}$ \cite{witt}.
The spinors are real
$\theta_{\alpha}=\theta_{\alpha}{}^{\dagger}$, $\alpha=1,2$, and
the adjoint of the  spinor is pure imaginary,
$\bar{\theta}=\theta^{\dagger}\gamma^{0}$. The charge conjugation
matrix, $C_{\alpha\beta}$, satisfies \be \ba{lll}
C^{-1}\gamma^{\mu}C=-\gamma^{\mu}{}^T\,,~~~&~~~C=-C^{T}=C^{-1}\,,~~~&~~~
C=\gamma^{0}\,. \ea \ee

Using the covariant derivatives given by \be \ba{ll} D_{\alpha} =
\displaystyle{\frac{\partial}{\partial{\bar{\theta}}^{\alpha}}
-i(\gamma^{\mu}\theta)_{\alpha}\frac{\partial}{\partial x^{\mu}}}
\,,~~~&~~~ {\bar{D}}^{\alpha} = D_{\beta}C^{-1\beta\alpha}=
\displaystyle{-\frac{\partial}
{\partial{\theta}_{\alpha}}+i(\bar{\theta}\gamma^{\mu})^{\alpha}\frac{\partial}
{\partial x^{\mu}}}\,, \ea \ee we consider the following action of
the $\N=1$ supersymmetric CP($N$) model \cite{golo,aref,aoya} in
three dimensions with a topological interaction ;
\begin{equation}
\S= \int dx^{3}d\theta^{2}
\Bigg[\frac{N}{2g}\Big({\bar{\nabla}}^{\alpha}
\bar{\Phi}_{i}\nabla_{\alpha}\Phi_{i} +
2\Sigma({\bar{\Phi}}_{i}\Phi_{i} - 1) \Big) -
i\textstyle{\frac{\kappa}{4}} N
\Big({\bar{F}}^{\alpha}\bar{\Phi}_{i} D_{\alpha}\Phi_{i} -
{\bar{D}}^{\alpha}\bar{\Phi}_{i}{\Phi_{i}}F_{\alpha}\Big)
\Bigg]\,. \label{cp}
\end{equation}
The first two terms corresponds to the supersymmetric  CP($N$)
model and the last two terms describes the topological interaction
\cite{itoh}. In the above action, $\Phi_{i}$,
$\bar{\Phi}_{i}=\Phi_{i}{}^{\dagger}$, $i=1,\cdots,N$ are complex
superfields, \be \Phi_{i}(x,\theta) = \phi_{i}(x) +
\bar{\theta}\psi_{i}(x) +
{\textstyle{\frac{1}{2}}}\bar{\theta}\theta F_{i}(x)\,, \ee and
$\Sigma$ is a real superfield serving as a Lagrange multiplier,
\be \Sigma(x,\theta) = \sigma(x) + \bar{\theta}\xi(x) +
{\textstyle{\frac{1}{2}}} \bar{\theta}\theta\alpha(x) \,. \ee The
gauge covariant derivatives are given by \be \ba{ll}
\nabla_{\alpha} = D_{\alpha} + iA_{\alpha}\,,~~&~~
\bar{\nabla}^{\alpha}=\bar{D}^{\alpha} - i\bar{A}^{\alpha}\,, \ea
\ee with a real spinor superfield, $A_{\alpha}$, \be A_{\alpha} =
\chi_{\alpha} + i(\gamma^{\mu}\theta)_{\alpha}A_{\mu} +
\theta_{\alpha}D +
{\textstyle{\frac{1}{2}}}\bar{\theta}\theta\omega_{\alpha} \,. \ee
$A_{\mu}$ is the usual  U(1) auxiliary gauge field and
$\bar{\nabla}^{\alpha}
\bar{\Phi}=(\nabla_{\beta}\Phi)^{\dagger}C^{-1\beta\alpha}$.

The above action is invariant under the U(1) gauge
transformation given by
\be \ba{ll} \Phi ~\rightarrow~
e^{i\Lambda}\Phi \,, ~~~&~~~ A_{\alpha} ~\rightarrow~
A_{\alpha} + D_{\alpha}\Lambda \,, \label{trans} \ea \ee
where
$\Lambda$ is a real scalar superfield,
\begin{equation}
\Lambda = b + \bar{\theta}f + {\textstyle{\frac{1}{2}}}\bar{\theta}\theta a \,.
\end{equation}
In the Wess-Zumino gauge, $f_{\alpha}=-\chi_{\alpha}$, $a=-D$,
\begin{equation}
A_{\alpha}(x,\theta) = i(\gamma^{\mu}\theta)_{\alpha}A_{\mu} +
{\textstyle{\frac{1}{2}}}\bar{\theta}\theta\omega_{\alpha} \,,
\end{equation}
and the real spinor superfield strength, $F_{\alpha}$,  takes the
form, \be F_{\alpha}=-{\bar{D}}^{\beta}D_{\alpha}A_{\beta} =
\omega_{\alpha}+(F_{\mu\nu}
\gamma^{\mu\nu}\theta)_{\alpha}-i\frac{1}{2}\bar{\theta}
\theta(\gamma^{\mu}\partial_{\mu}\omega)_{\alpha} \,, \ee where
$F_{\mu\nu}=\partial_{\mu}A_{\nu}-\partial_{\nu}A_{\mu}$. Note
that $\bar{F}^{\alpha}=F_{\beta}C^{-1\beta\alpha}$ and
$F_{\alpha}$ is gauge invariant due to the identity,
${\bar{D}}^{\beta}D_{\alpha}D_{\beta}=0$. The action is also
invariant under the global SU($N$) symmetry, $\Phi_{i}\rightarrow
U_{i}{}^{j}\Phi_{j}$. \\

Essentially the above action is the supersymmetric generalization of
the bosonic model one of the authors studied previously
\cite{itoh}. It is  demonstrated in the  Appendix that the spinor
superfield, $A_{\alpha}$,
 can be eliminated from the action, (\ref{cp}), using the equation of motion as in the
case of the bosonic model, and the resulting action corresponds
to the supersymmetric generalization of the higher derivative
CP($N$) model with the Wess-Zumino-Witten term.
%%%%%%%%%%%%%%%%%%%%%%%%%%%%%%%%%%%%%%%%%%%%%%%%%%%%%%%%%%%%%%%%%%%%%%%%%%%%%%%%%
%%%%%%%%%%%%%%%%%%%%%%%%%%%%%%%%%%%%%%%%%%%%%%%%%%%%%%%%%%%%%%%%%%%%%%%%%%%%%%%%%
%%%%%%%%%%%%%%%%%%%%%%%%%%%%%%%%%%%%%%%%%%%%%%%%%%%%%%%%%%%%%%%%%%%%%%%%%%%%%%%%%
\subsection{The effective action and the gap equation}
In order to obtain the effective action and gap equation, we
integrate out the complex scalar superfield, $\Phi$. We first
rewrite the action (\ref{cp}) in the Gaussian form as
\begin{equation}
\S= \int {d}z^{5} \left[{\textstyle{\frac{N}{2g}}}{\bar{\Phi}}_{i}
\Delta\Phi_{i} - {\textstyle{\frac{N}{g}}}\Sigma\right] \,,
\end{equation}
with a differential operator, $\Delta$,
\begin{equation}
\Delta = -\bar{D}D - i(\bar{D}A) - 2i\bar{A}D + \bar{A}A - i\kappa g \bar{F}D
+ 2\Sigma \,.
\label{pro}
\end{equation}
Introducing the external source $J_{i}$ for ${\bar\Phi_{i}}$ (
and  similarly for ${\bar{J}}_{i}$ for ${\Phi}_{i}$) given by
\begin{equation}
J_{i} = n_{i} + \bar{\theta}{\eta}_{i}+
{\textstyle{\frac{1}{2}}}\bar{\theta}\theta{j}_{i} \,,
\end{equation}
we obtain the generating functional,
\begin{equation}
Z[J_{i},{\bar{J}}_{i}] = \int {\cal D}\Phi_{i}{\cal D}{\bar\Phi}_{i}
{\cal D}A{\cal D}\bar{A}{\cal D}\Sigma
~{\rm{exp}}\Bigg[i\int {d}z^{5}\left({\textstyle{\frac{N}{2g}}}
{\bar{\Phi}}_{i}\Delta\Phi_{i} - {\textstyle{\frac{N}{g}}}\Sigma
+ {\textstyle{\frac{1}{2}}}{\bar{J}}_{i}\Phi_{i}
+ {\textstyle{\frac{1}{2}}}{\bar{\Phi}}_{i}J_{i} \right)\Bigg] \,.
\label{gen}
\end{equation}
Integrating out the scalar superfields $\Phi_{i}$,
${\bar{\Phi}}_{i}$ yields
\be \ba{c} Z[J_{i} \, , \, {\bar{J}}_{i}]
= \int {\cal D}A{\cal D}\bar{A}{\cal D}\Sigma
\,\,\, e^{iW_{\rm eff}} \,,\\
{}\\
W_{\rm eff} = iN``\mbox{Str}(\ln \Delta)" - \int {d}z^{5}\, {\textstyle{\frac{N}
{g}}}\Sigma - {\textstyle{\frac{g}{2N}}}\int {d}z^{5}\,
{\bar{J}}_{i}{\Delta}^{-1}J_{i} \,,
\ea
\label{w}
\ee
where $``\mbox{Str}(\ln \Delta)"$ can be explicitly read off from (\ref{str}),
as done in the next subsection \ref{s1cp}. \\
Here we  approximate (\ref{w}) by  the large $N$ saddle point
method,
%and expand it around the classical solutions, $\Phi_{i}$ and $\bar{\Phi}_{i}$.
%of (\ref{gen}) enables us to express the external sources in
%terms of the vacuum expectation values of $\Phi_{i}$ and
%$\bar{\Phi}_{i}$ in the large $N$ saddle point method, \be \ba{ll}
%J_{i}=-\frac{N}{g}\Delta{\Phi_{i}}^{c} \,, ~~&~~
%\bar{J}_{i}=J_{i}{}^{\dagger} \,, \ea \ee where ${\Phi_{i}}^{c}$
%is the  classical superfield.
and taking the Legendre transformation of $W_{\rm eff}$, as usual,
we get the effective action in the leading order, \be
\S_{\rm{eff}}({\Phi_{i}}, {{\bar{\Phi}}_{i}}, \Sigma) = iN
``\mbox{Str}(\ln \Delta)" + \int {d}z^{5}
{\textstyle{\frac{N}{2g}}}
\left({{\bar{\Phi}}_{i}}\Delta{\Phi_{i}} - 2\Sigma \right) \,.
\label{eff} \ee

Let us turn to the gap equation which can be derived from the
stationary  conditions of the effective potential. Taking
advantage of the global U($N$) symmetry, we write expectation
values of $\Phi_{i}$ and $\bar{\Phi}_{i}$ as \be
(\langle\Phi_{1}\rangle,\langle\Phi_{2}\rangle,\cdots,\langle\Phi_{N}\rangle)
=(0,0,\cdots,\sqrt{g}V) \,, \ee and $\langle\Sigma\rangle$ for
that of $\Sigma$. Turning off the spinor superfield,
$A_{\alpha}=0$, we obtain the effective potential \be \ba{ll}
U_{\rm{eff}}&\! \equiv -{\Omega}^{-1}\S_{\rm{eff}} \\
{}&{}\\
{}& = -N\int {d}\theta^{2} \,
( \bar{V}V - {\textstyle{\frac{1}{g}}} ) \langle{\Sigma}\rangle
- i{\textstyle{\frac{1}{2}}}N \int {d}\theta^{2}
\big{\langle} x | \big{\lbrace}{\bar{\theta}}^{\alpha} \, , \, \big{\lbrack}
\theta_{\alpha} \, , \, \rm{ln}( -\bar{D}D + 2\langle{\Sigma}\rangle )
\big{\rbrack}\big{\rbrace} | x \big{\rangle} \\
{}&{}\\
{}& = -N\int {d}\theta^{2} \, ( \bar{V}V -
{\textstyle{\frac{1}{g}}} ) \langle{\Sigma}\rangle - iN\int
{d}\theta^{2} \, \displaystyle{\lim_{z^{\prime}\rightarrow z}} \,
\langle z^{\prime} | \ln\big(-\bar{D}D + 2\langle{\Sigma}\rangle
\big)|z\rangle \,, \ea \label{ep}\ee
where $\Omega$ is the three-dimensional spacetime volume.
%Note that the second term of
%each line is independent of the spacetime coordinate, $x$.

The stationary conditions for the effective potential read \be
\ba{ll}
\bar{V}\langle{\Sigma}\rangle = V\langle{\Sigma}\rangle = 0 \,, \\
{}&{}\\
\displaystyle{\bar{V}V - \frac{1}{g}} + \displaystyle{
{\frac{i}{(2\pi)^{3}}\int dk^{3}~
\frac{1}{~{k}^{2}-{\langle\Sigma\rangle}^{2}+i\varepsilon}= 0}} \,.
\ea \label{sc} \ee The UV divergence appears in the second
formula and with the momentum cut-off  $\Lambda$,
it becomes \be \bar{V}V - \frac{1}{g} + \frac{\Lambda}{2{\pi}^{2}}
- \frac{\langle\Sigma\rangle} {4\pi} = 0 \,. \label{gpe} \ee
Now we  introduce an arbitrary  scale parameter,
$\mu$, and a renormalized coupling constant, $g_{r}$,  which satisfy
\be \bar{V}V - \frac{1}{g_{r}} + \frac{\mu}{2{\pi}^{2}} -
\frac{\langle\Sigma\rangle} {4\pi} = 0 \,. \label{gpe2} \ee Then,
the renormalized coupling constant and the bare coupling constant are
related by \be \frac{1}{g} = \frac{1}{g_{r}} +
\frac{\Lambda}{2{\pi}^{2}} - \frac{\mu}{2{\pi}^{2}} \,.
\label{gpe3} \ee With the introduction of the dimensionless
coupling $u$ defined by $u \equiv \Lambda g$, we obtain the
$\beta$-function as \be \beta(u) \equiv
\Lambda\frac{du}{d\Lambda} = u\Big(1-\frac{u} {2{\pi}^{2}}\Big)
\,. \label{bf} \ee  This $\beta$-function shows a nontrivial UV
fixed point at $u =2 \pi^2\equiv u^{\ast}$. In the continuum limit,
$\Lambda \rightarrow \infty$, we have $u \rightarrow u^{\ast}$.\\

From (\ref{sc}) and (\ref{bf}), we find that there are
two phases : \\
(i) SU($N$) symmetric phase for $u > 2{\pi}^{2}$,
\be
\ba{ll}
V=\bar{V} = 0 \,, \\
{}&{}\\
\langle{\Sigma}\rangle = \langle{\sigma}\rangle + \bar{\theta}\langle{\xi}
\rangle + {\textstyle{\frac{1}{2}}}\bar{\theta}\theta\langle{\alpha}\rangle
\equiv m
%= 4\pi\mu\Big(\frac{1}{2{\pi}^{2}}- \frac{1}{u_{r}}\Big)
\,.
\ea
\label{sp}
\ee
\\
(ii) SU($N$) $\rightarrow$ SU($N-1$)$ \times$ U($1$)  broken phase
for $u < 2{\pi}^{2}$, \be \ba{l} V =
{\textstyle{\frac{1}{\sqrt{g}}}}\left[ \langle{\phi_{N}}\rangle +
\bar{\theta}\langle{\psi_{N}}\rangle +
{\textstyle{\frac{1}{2}}}\bar{\theta}
\theta\langle{F_{N}}\rangle \right] \equiv v \ne 0 \,, \\
{}\\
\bar{V} = {\textstyle{\frac{1}{\sqrt{g}}}}\left[ \langle{\bar{\phi}}_{N}\rangle
+ \langle{\bar{\psi}}_{N}\rangle \theta + {\textstyle{\frac{1}{2}}}
\bar{\theta}\theta\langle\bar{F}_{N}\rangle \right] \equiv \bar{v} \ne 0 \,, \\
{}\\

\bar{V}V = \Lambda\Big(\frac{1}{u} - \frac{1}{2{\pi}^{2}}\Big)\,,\\
{}\\
\langle{\Sigma}\rangle = 0\,. \ea \label{bp} \ee Note that in
both phases, from $\langle{\xi}\rangle =
\langle{{\bar{\psi}}_{N}} \rangle = \langle{\psi_{N}}\rangle = 0$
and $\langle{\alpha}\rangle = \langle{{\bar{F}}_{N}} \rangle =
\langle{F_{N}}\rangle = 0$, the supersymmetry is unbroken. For the
symmetric phase, Eq.(\ref{gpe}) becomes \be \Big{(}\frac{1}{u} -
\frac{1}{u^{\ast}}\Big{)}\Lambda + \frac{m}{4\pi} = 0 \,.
\label{gap} \ee This gap equation is identical to that in the
bosonic theory \cite{aref,itoh}, and the mass scale, $m$, is an
cutoff independent  parameter of the theory.

% independent
%of the ultraviolet cutoff, $\Lambda$, by making $\Lambda$ depends
%on the coupling $u$. So the scale invariance condition $\Lambda d
%m /d\Lambda=0$ gives the $\beta$-function, \be \beta(u) =
%u\Big{(}1 - \frac{u}{u^{\ast}}\Big{)} \,, \label{be} \ee which
%From (\ref{ge}) and due to the fact of the mass, $m$, being scale
%invariant, the relation of renormalized dimensionless coupling
%$u_{r}$ at a reference energy scale $\mu$ and dimensionless
%coupling $u$ is given by \be \Big{(}\frac{1}{u} -
%\frac{1}{u^{\ast}}\Big{)}\Lambda = \Big{(}\frac{1}{u_{r}} -
%\frac{1}{{u_{r}}^{\ast}}\Big{)}\mu \,. \label{gap} \ee

\newpage
%%%%%%%%%%%%%%%%%%%%%%%%%%%%%%%%%%%%%%%%%%%%%%%%%%%%%%%%%%%%%%%%%%%%%%%%%%%%%%%%%
%%%%%%%%%%%%%%%%%%%%%%%%%%%%%%%%%%%%%%%%%%%%%%%%%%%%%%%%%%%%%%%%%%%%%%%%%%%%%%%%%
%%%%%%%%%%%%%%%%%%%%%%%%%%%%%%%%%%%%%%%%%%%%%%%%%%%%%%%%%%%%%%%%%%%%%%%%%%%%%%%%
\subsection{One loop effective action and renormalization\label{s1cp}}
In this subsection, we consider the symmetric phase and  redefine
$\Sigma$ as
 \be \Sigma = {m + {\Sigma}^{\prime}}\,. \label{sh}
\ee After some manipulation for the one loop effective action
using the superfield formalism developed in section
\ref{general},  we obtain the following large $N$ effective action up to the
quadratic terms,
 \be \S_{\rm{eff}} = N\! \int \! {d}z^{5} \left[ \ba{ll}
{\textstyle{\frac{1}{2g}}}\bar{\Phi}\Big{(}
-\bar{D}D+2m-i(\bar{D}A)-2i\bar{A}D
+\bar{A}A-i\kappa g\bar{F}D+2{\Sigma}^{\prime}\Big{)}\Phi \\
{}\\
-\Big{(}\frac{1}{u}-\frac{1}{u^{\ast}}\Big{)}\Lambda(m{\delta}^{(2)}(\theta)
+{\Sigma}^{\prime}) -{\textstyle{\frac{m}{4\pi}}}{\Sigma}^{\prime}
- {\textstyle{\frac{m^{2}}{8\pi}}}{\delta}^{(2)}(\theta) \\
{}\\
- {\textstyle{\frac{1}{2}}}{\Sigma}^{\prime}{\Pi}_{{\Sigma}^{\prime}}
(-i\partial){\Sigma}^{\prime}
- {\textstyle{\frac{1}{16}}}\bar{F}{\Pi}_{1}(-i\partial)F
+ {\textstyle{\frac{m}{4}}}\bar{A}{\Pi}_{1}(-i\partial)F \\
{}\\
- {\textstyle{\frac{1}{2}}}\kappa g\bar{A}{\Pi}_{2}(-i\partial)F
- {\textstyle{\frac{1}{8}}}(\kappa g)^{2}\bar{F}{\Pi}_{2}(-i\partial)F
\ea\right] \,,
\label{ef}
\ee
where
\be
\ba{ll}
{\Pi}_{{\Sigma}^{\prime}}(p) = \frac{\bar{D}(p,\theta)D(p,\theta)+4m}
{8\pi\sqrt{-p^{2}}}~{\rm arctan}\Big{(}\frac{\sqrt{-p^{2}}}{2m}\Big{)} \,, \\
{}&{}\\
{\Pi}_{1}(p) = \frac{1}{4\pi\sqrt{-p^{2}}}~{\rm arctan}
\Big{(}\frac{\sqrt{-p^{2}}}{2m}\Big{)} \,, \\
{}&{}\\
{\Pi}_{2}(p) = \frac{\Lambda}{2\pi^{2}} - \frac{m}{4\pi} +
\frac{p^{2}-2mp\!\!\!/}{8\pi\sqrt{-p^{2}}}~{\rm arctan}
\Big{(}\frac{\sqrt{-p^{2}}}{2m}\Big{)}. \ea \ee Note that  in the
above action,  the terms containing ${\delta}^{(2)}(\theta)$ come
from the effective potential (\ref{ep}), and the $\Lambda$
dependent term in the second line becomes finite through the gap
equation (\ref{gap}). Also, the derivative expansion in the third
and fourth lines are manifest supersymmetric because they are all
given in term of $(\bar D D)^2= -4\partial_\mu\partial^\mu=4p^2$.

As in the bosonic case \cite{itoh}, the above action is
renormalizable in spite of the linear divergence term.
Introducing the $Z$ factor for the coupling constant
 \be
 \ba{cc}
Z^{-1} \equiv\displaystyle{ \frac{g_r}{g} = 1+
\frac{u_r}{{u}^{\ast}}\left(\frac{\Lambda}{\mu}-1\right)}
\,,~~~&~u_r\equiv \mu g_r\,,\ea \ee and redefining
$\Phi\equiv\sqrt{Z}\Phi_{r}$ in order to cancel the $Z$ factor
from the coupling constant renormalization, we recast the kinetic
term as  \be \frac{1}{g}\bar{\Phi}(-\bar{D}D+2m)\Phi =
\frac{1}{g_{r}}{\bar{\Phi}}_{r} (-\bar{D}D+2m)\Phi_{r} \,. \ee
 This makes sure the  kinetic term is   UV finite by itself.
The spinor superfield, $A_\alpha$, and  the real superfield,
$\Sigma^\prime$, do not need any wave function renormalization.
The induced supersymmetric Maxwell-Chern-Simons term  contains a
linear divergence in $\Pi_2(p)$. However,  the $\beta$-function
(\ref{bf}) tells us that the dimensionless coupling $u\equiv
g\Lambda$ goes to the UV fixed point, $u^{\ast}=2\pi^{2}$, as the
cutoff $\Lambda$ goes to the infinity. Therefore,  in the
continuum limit, $\Lambda \rightarrow \infty$, we obtain the UV
finite result, $g{\Pi}_{2}(p) \rightarrow 1$,
$g^{2}{\Pi}_{2}(p)\rightarrow 0$, including  the supersymmetric
Maxwell-Chern-Simons terms, \be -
{\textstyle{\frac{1}{16}}}\bar{F}{\Pi}_{1}(-i\partial)F +
{\textstyle{\frac{m}{4}}}\bar{A}{\Pi}_{1}(-i\partial)F -
{\textstyle{\frac{1}{2}}}\kappa \bar{A}F \,. \ee After all, the
one loop effective action (\ref{ef}) leads, in the continuum
limit, to a scalar superfield theory coupled to the derivatively
expanded  Maxwell-Chern-Simons theory;
 \be
\S_{\rm{eff}} = N\! \int \! {d}z^{5} \left[ \ba{l}
{\textstyle{\frac{1}{2g_r}}}{\bar\Phi}_r\Big{(}
-\bar{D}D+2m-i(\bar{D}A)-2i\bar{A}D
+\bar{A}A-i\kappa g_r\bar{F}D+2{\Sigma}^{\prime}\Big{)}\Phi_r \\
{}\\
+ {\textstyle{\frac{m^{2}}{8\pi}}}{\delta}^{(2)}(\theta) -
{\textstyle{\frac{1}{2}}}{\Sigma}^{\prime}{\Pi}_{{\Sigma}^{\prime}}
(-i\partial){\Sigma}^{\prime} -
{\textstyle{\frac{1}{16}}}\bar{F}{\Pi}_{1}(-i\partial)F +
{\textstyle{\frac{m}{4}}}\bar{A}\left({\Pi}_{1}(-i\partial)-
\frac{2\kappa}{m}\right)F
 \ea\right] \,. \label{ef1}
 \ee

At the lowest order of the expansion in $\partial/m$,  the
supersymmetric Maxwell-Chern-Simons theory is induced with the
coefficient of the Chern-Simons term given by $ \frac{1}{32\pi}
-\frac{1}{2} \kappa$. The origin of the parity-violating
Chern-Simons term with the coefficient, $\frac{1}{32\pi}$, is due
to the parity violating fermion mass term in the first line of
the action (\ref{ef}) \cite{dese}, while  the topological
interaction term shifts the coefficient by an amount of
$-\frac{1}{2} \kappa$. It is interesting to note that the next
order of the expansion of the Chern-Simons term yields the
supersymmetric generalization of the higher derivative
Chern-Simons extensions, $ \sim m^{-2}\bar{A}(\bar D D)^2 F$
\cite{jack}, and with the choice of the coefficient $
\kappa\equiv\frac{1}{16\pi}$, the gauge sector of the effective
action becomes the supersymmetric generalization of the extended
topological massive electrodynamics model of Ref.~\cite{jack}

%Hence we can
%conclude that the large $N$ effective action in our model is
%renormalizable and the gauge sector is equivalent to the
%supersymmetric Maxwell-Chern-Simons theory.

%%%%%%%%%%%%%%%%%%%%%%%%%%%%%%%%%%%%%%%%%%%%%%%%%%%%%%%%%%%%%%%%%%%%%%%%%%%%%%%%%%%%%%%%%%%%%%%%%%%%%%%%%%%%%%%%%%%%%%%%%%%%%%%%%%%%%%%%%%%%%%%%%%%%%%
%%%%%%%%%%%%%%%%%%%%%%%%%%%%%%%%%%%%%%%%%%%%%%%%%%%%%%%%%%%%%%%%%%%%%%%%%%%%%%%%%%%%%%%%%%%%%%%%%%%%%%%%%%
%%%\section{Remarks\label{secCon}}
%%%%%%

\acknowledgments{PO was supported by Korea Research Foundation
grant (KRF-2002-042-C00010). JHP's work is the result of research
activities, Astrophysical Research Center for the Structure and
Evolution of the Cosmos, supported by Korea Science and
Engineering Foundation. JHP was also partly supported
by the CNNC. } %%%%%%%%%%%%%%%%%%%%%%%%%%%%%%%%%%%%%%%%%%%%%%%%%%%%%%%%%%%%%%%%%%%%%%%%%%%%%%%%%%%%%%%%%%%%%%%%%%%%%%%%%%%%%%%%%%%%%%%%%%%%%
%%%%%%%%%%%%%%%%%%%%%%%%%%%%%%%%%%%%%%%%%%%%%%%%%%%%%%%%%%%%%%%%%%%%%%%%%%%%%%%%%%%%%%%%%%%%%%%%%%%%%%%%%%%%%%%%%%%%%%%%%%%%%
%%%%%%%%%%%%%%%%%%%%%%%%%%%%%%%%%%%%%%%%%%%%%%%%%%%%%%%%%%%%%%%%%%%%%%%%%%%%%%%%%%%%%%%%%%%%%%%%%%%%%%%%%%%%%%%%%%%%%%%%%%%%%
%%%%%%%%%%%%%%%%%%%%%%%%%%%%%%%%%%%%%%%%%%%%%%%%%%%%%%%%%%%%%%%%%%%%%%%%%%%%%%%%%%%%%%%%%%%%%%%%%%%%%%%%%%%%%%%%%%%%%%%%%%%%%
\newpage
\appendix
%%%
%%\begin{center}
%%\large{\textbf{Appendix}}
%%\end{center}
%%\setcounter{equation}{0}
%%\renewcommand{\theequation}{A.\arabic{equation}}
%%%
\section{Appendix}
Here we show that the spinor superfield, $A_{\alpha}$,
 can be eliminated from the action, (\ref{cp}), using the equation of motion, and the resulting
action corresponds to the supersymmetric generalization of the higher derivative  CP($N$)
model \cite{itoh}.

The equation of motion for the spinor superfield reads from
\be
\delta_{A}\S=\displaystyle{\int d^3x d^2\theta} \frac{N}{2g} \left[\delta
{\bar{A}}^\alpha\left(A_\alpha -i\Phi^\dagger D_\alpha\Phi-
\frac{g\kappa}{2}{\bar{D}}^\beta D_\alpha(i\Phi^\dagger D_\beta\Phi)\right)+c.c.\right]\,,
\ee
so that
\begin{equation}
A_\alpha=J_\alpha+\frac{g\kappa}{2}t_\alpha \,,
\end{equation}
where   $J_\alpha=i\Phi^\dagger D_\alpha\Phi$, $t_\alpha={\bar{D}}^\beta
D_\alpha J_\beta$ are  the current density and  the topological current density respectively,
satisfying ${\bar{D}}^\alpha t_\alpha=0$.

Substituting the expression into the action we obtain
\begin{equation}
\S^{\prime}=\frac{N}{2g}\int d^3 x d^2 \theta \left(
\bar{D}^\alpha\Phi^\dagger D_\alpha\Phi -{\bar{J}}^\alpha
J_\alpha -\frac{g^2{\kappa}^2}{4} {\bar{t}}^\alpha t_\alpha
\right) +\S_{WZW} \,.
\end{equation}
The third term inside the bracket is the supersymmetric generalization of the fourth
order derivative term \cite{itoh}, while $\S_{WZW}$ is the
supersymmetric  Wess-Zumino-Witten term in three dimensions \cite{ferr},
\begin{equation}
\S_{WZW}=-\frac{\kappa}{4}N\int d^3xd^2\theta\left(
{\bar{t}}^\alpha J_\alpha +{\bar{J}}^\alpha t_\alpha\right)\,.
\end{equation}

\end{document}